\documentclass[aps,prb,showpacs,twocolumn,floatfix,letterpaper]{revtex4-1}

\usepackage[intlimits,sumlimits]{amsmath}
\usepackage{amsfonts,amssymb}
\usepackage{bm}
\usepackage{graphicx}
\usepackage[colorlinks,allcolors=blue]{hyperref}
\usepackage{cases}

\begin{document}
\newcommand\UCSD{Department of Physics, University of California San Diego, 9500 Gilman Drive, La Jolla, California 92093, USA}

\title{Electronic response of graphene to linelike charge perturbations}

\author{B.-Y. Jiang and M.~M.~Fogler}
\affiliation{\UCSD}

\date{\today}

\begin{abstract}

The problem of electrostatic screening of
a charged line by undoped or weakly doped graphene is
treated beyond the linear-response theory.
The induced electron density is found to be approximately doping independent,
$n(x) \sim x^{-2} \log^2 x$, at intermediate distances $x$ from the charged line.
At larger $x$, twin $p$-$n$ junctions may form if the external perturbation is repulsive for graphene charge carriers.
The effect of such inhomogeneities on
conductance and quantum capacitance of graphene is calculated.
The results are relevant for
transport properties of graphene grain boundaries and for local electrostatic control of graphene with ultrathin gates.

\end{abstract}
\maketitle

\section{Introduction}
\label{sec:intro}

\begin{figure}[b]
\begin{center}
\includegraphics[width=2.1in]{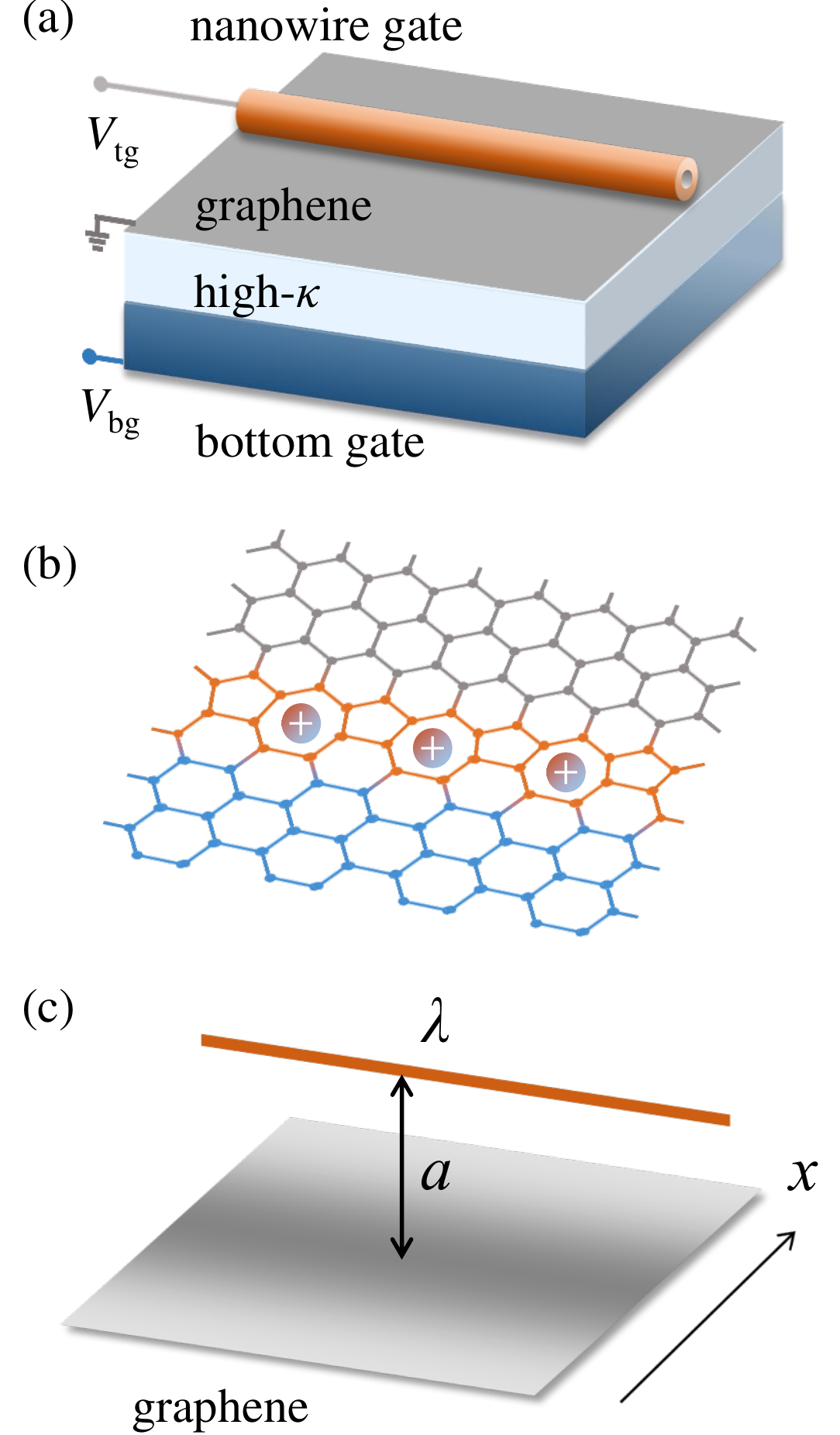}\\
\end{center}
\caption{(Color online) Models of linear charged perturbations in graphene devices. 
(a) Nanowire top gate. The bottom gate separated from graphene by an insulator of high dielectric constant $\kappa$ may be useful for additional control.
(b) Grain boundary (pentagon-heptagon chain) with charged adsorbates (circles).
(c) Charged string off the graphene plane.}
\label{fig:schematic}
\end{figure}

One of the properties that make graphene an attractive platform for electronic devices is tunability of its charge carrier density through electrostatic gating.
The gates can be brought into immediate proximity of graphene,
which enables one to control doping of this material very efficiently.
Local gating on ultrasmall lengthscales is attainable by utilizing nanowire~\cite{Young2009qik} or nanotube~\cite{Ni2015xxx} gates [Fig.~\ref{fig:schematic}(a)].
Physical insight into fundamental characteristics of such devices can be gained from a simplified problem of how graphene
responds to a linelike external charge [Fig.~\ref{fig:schematic}(c)].
This problem is also relevant for understanding properties of
grain-boundary defects [Fig.~\ref{fig:schematic}(b)] in graphene grown by chemical-vapor deposition.~\cite{Biro2013gbg,Cummings2014ctp,Yazyev2014pgo}

The problem of
screening of a linelike charge by electrons in graphene is an interesting
challenge because
the usual linear-response theory~\cite{Radchenko2013ecl} fails when the line is highly charged and/or when graphene is lightly doped.
Previous studies of one-dimensional charge perturbations in graphene eschewed solving this difficult problem 
resorting instead to \textit{ad hoc} approximations for the induced carrier density profile.~\cite{Young2009qik, Fei2013epp}
Accurate determination of this profile requires numerical calculations, e.g.,
finding the self-consistent solution of
the Dirac equation for electron wavefunctions
and the Poisson equation for the electrostatic potential $\Phi(x)$ as a function of the in-plane coordinate $x$ transverse to the charged line.
However, if $\Phi(x)$ varies smoothly on the scale of the local Fermi length $k_F^{-1}(x)$,
a simpler approach based on the
Thomas-Fermi approximation (TFA)
can be applied.~\cite{Landau1981qmn}
We show that approximate solutions of the TFA equations for the electron density can be derived in certain limits.
Using these analytical solutions and numerical simulations,
we compute two other important observables
amenable to experimental probes: the conductance and the gate capacitance.

Let us introduce our key notations and assumptions.
We denote the linear charge density of the external perturbation by $e\lambda$.
Without loss of generality, we take $\lambda$ to be positive
(unless specified otherwise).
We assume that the unperturbed electron density $n_\infty$ of graphene is uniform.
To distinguish between $n$-type and $p$-type doping,
we define the Fermi momentum corresponding to $n_\infty$ as a signed quantity,
\begin{equation}
k_\infty = \mathrm{sgn}(n_\infty) |\pi n_\infty|^{1 / 2}.
\label{eqn:k_infty}
\end{equation}
We assume that
the external charge $e n_\mathrm{ext}(x)$ has a Lorentzian density distribution,
\begin{equation}
	n_\mathrm{ext}(x)
	 = \frac{\lambda}{\pi}\frac{a}{x^2 + a^2}\,.
\label{eqn:n_ext}
\end{equation}
The actual profile of the external charge may of course be somewhat different.
For example, the charge distribution of a grain boundary probably does not have power-law tails.
However, the role of parameter $a$ in Eq.~\eqref{eqn:n_ext} is mainly to regularize the response of graphene at very short distances,~\cite{Fogler2007shc} i.e., it serves as a short-distance cutoff.
As long as we are not interested in microscopic physics at $|x| \lesssim a$,
Eq.~\eqref{eqn:n_ext} can be adopted as a convenient model.
In all examples mentioned above (nanowire and nanotube gates and also grain boundaries in graphene) realistically achievable $a$ can be as small as a few nanometers.
The particular functional form of Eq.~\eqref{eqn:n_ext} corresponds to the idealized model shown in Fig.~\ref{fig:schematic}(c) where the external charge is located off the graphene plane and is truly one-dimensional.
This can be seen from the fact that the electrostatic potential created in the graphene plane by the out-of-plane charged line is equal to that created by the \textit{in-plane} Lorentzian charge distribution~\eqref{eqn:n_ext}:
\begin{align}
\Phi_\mathrm{ext}(x) & = \frac{e}{\kappa}
\int_{-\infty}^{\infty} d x'
n_\mathrm{ext}(x') \log \frac{1}{(x - x')^2}
\label{eqn:Phi_ext}\\
&= \frac{e\lambda}{\kappa}
 \log \frac{1}{x^2 + a^2}\,.
\label{eqn:Phi_ext_II}
\end{align}
In general, the effective width parameter $a$ should be taken as the larger of the actual physical width of the charge distribution and its distance to the graphene plane.
In this article we assume that electron-electron interaction in graphene is weak, i.e., we consider the dimensionless coupling constant $\alpha = e^2 / \kappa \hbar v$ a small parameter in the problem.~\cite{Kotov2012eei}
Here $v = 10^8\,\mathrm{cm}\,\mathrm{s}^{-1}$ is the graphene Fermi velocity. 
By choosing a substrate with a large dielectric constant $\kappa$ [Fig.~\ref{fig:schematic}(a)], it is possible to make $\alpha \simeq 2.2/\kappa$ significantly smaller than unity.
However, it is difficult to make $\alpha$ truly small, so
as a rule we do not treat $\log \alpha$ as a small parameter.
Finally, we assume that the graphene is not too highly doped,
\begin{equation}
	|k_\infty| \ll \frac{1}{\alpha a}\,.
\label{eqn:doping}
\end{equation}
\begin{figure}
\includegraphics[width=3.0in]{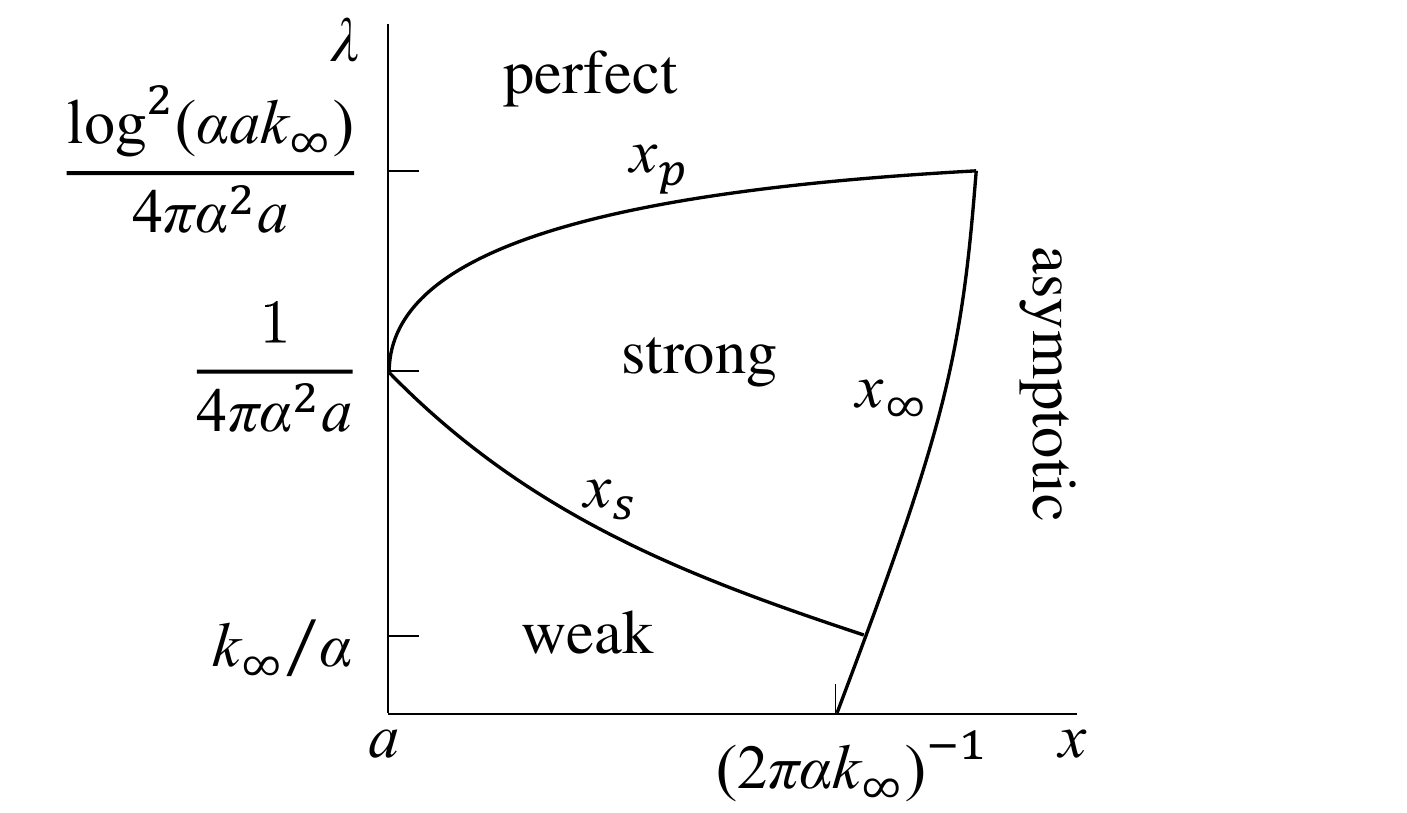}
\caption{Schematic diagram of screening regimes. 
In the `weak' region screening of the external potential is poor;
the response is linear [Eq.~\eqref{eqn:n_ind_ln_weak}] at $\lambda <  k_\infty / \alpha$  and nonlinear [Eq.~\eqref{eqn:n_ind_nl}] at
larger $\lambda$.
In the `strong' regime the external potential is greatly reduced and the induced density profile is given by Eq.~\eqref{eqn:n_ind_nl}.
In the `asymptotic' regime the density profile follows Eq.~\eqref{eqn:n_ind_asymp}.
In the `perfect' regime graphene maintains local charge neutrality,
apart from small corrections [Eq.~\eqref{eqn:n_ind_np}].
This diagram is drawn assuming graphene is not too heavily doped,
Eq.~\eqref{eqn:doping};
otherwise, the `weak' and `strong' screening regions would disappear.
}
\label{fig:regimes}
\end{figure}

Depending on the relation between $\lambda$, $k_\infty$, and $x$, the response of graphene can be either weak or strong and either linear or nonlinear (Fig.~\ref{fig:regimes}).
The degree of nonlinearity is controlled by the dimensionless parameter
\begin{equation}
\tilde{\lambda} = \frac{\alpha \lambda}{k_\infty}\,.
\label{eqn:tilde_lambda}
\end{equation}
Linear screening is realized if $\tilde{\lambda} \ll 1$ (the bottom part of the `weak' region in Fig.~\ref{fig:regimes}) where the induced electron density
\begin{equation}
n_\mathrm{ind}(x) \equiv n(x) - n_\infty
\label{eqn:n_ind_def}
\end{equation} 
scales linearly with $\lambda$.
On the other hand,
if $\tilde{\lambda} > 1$, the response is nonlinear.
A conspicuous manifestation of the nonlinearity is found in the region labeled `strong' in Fig.~\ref{fig:regimes},
where the induced density can be approximated by a `universal' (doping-independent) form $n_\mathrm{ind}(x) \sim x^{-2}\log^2 x$.
In the large-$x$ `asymptotic' regime of Fig.~\ref{fig:regimes}, the induced density exhibit a power-law decay
\begin{equation}
 n_\mathrm{ind}(x) \simeq \frac{b}{x^2}\,,
\label{eqn:n_ind_asymp}
\end{equation} 
where the dependence of $b$ on $\lambda$ is linear
[Eq.~\eqref{eqn:n_ind_ln} of Sec.~\ref{sec:linear}]
if $\tilde{\lambda}$ is small and logarithmic
if $\tilde{\lambda}$ is large [Eq.~\eqref{eqn:x_infty}].
The `perfect' screening region is included in Fig.~\ref{fig:regimes} for completeness.
Here graphene maintains charge neutrality locally,
i.e., the induced density
is close to the external one,
$n_\mathrm{ind}(x) \simeq n_\mathrm{ext}(x)$, 
so that the model assumption~\eqref{eqn:n_ext} must be critically revisited.

The crossovers among the predicted screening regimes can be
systematically studied in experiments using devices that have both top
and bottom gates, Fig.~\ref{fig:schematic}(a).
High linear charge densities with $\alpha\lambda \sim (1\,\mathrm{nm})^{-1}$ are quite feasible to achieve with top gate voltages $V \sim 1\,\mathrm{V}$.
For lightly doped graphene, $n_\infty = 10^{11}\,\mathrm{cm}^{-2}$, the corresponding $\tilde{\lambda} \sim 20$ is deep in the nonlinear regime.

In the remainder of this article we derive detailed formulas for the carrier density profiles, verify them by numerical simulations, then make predictions for capacitance and transport measurements.

\section{Linear response}
\label{sec:linear}

\begin{figure*}
	\begin{center}
		\includegraphics[height=2in]{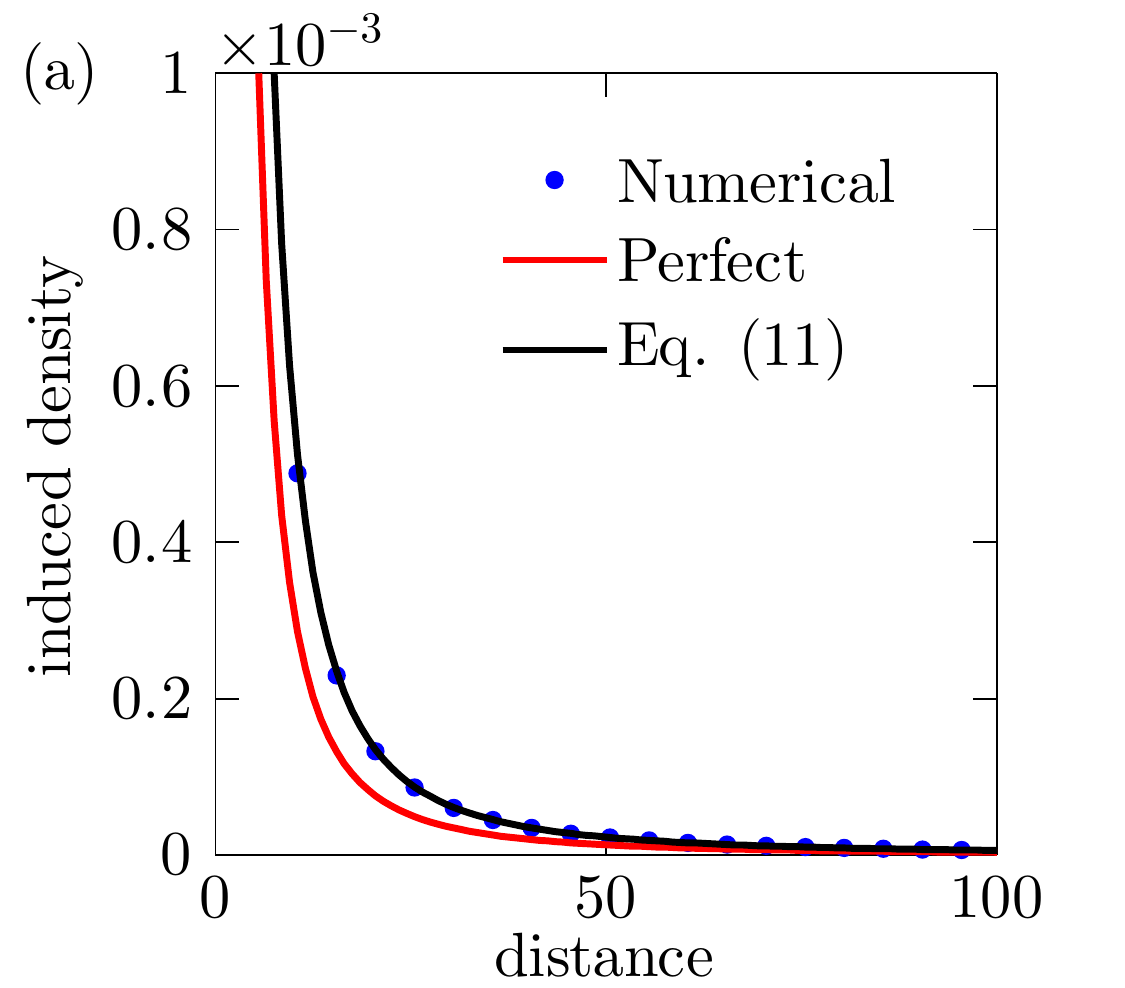}	
		\includegraphics[height=2in]{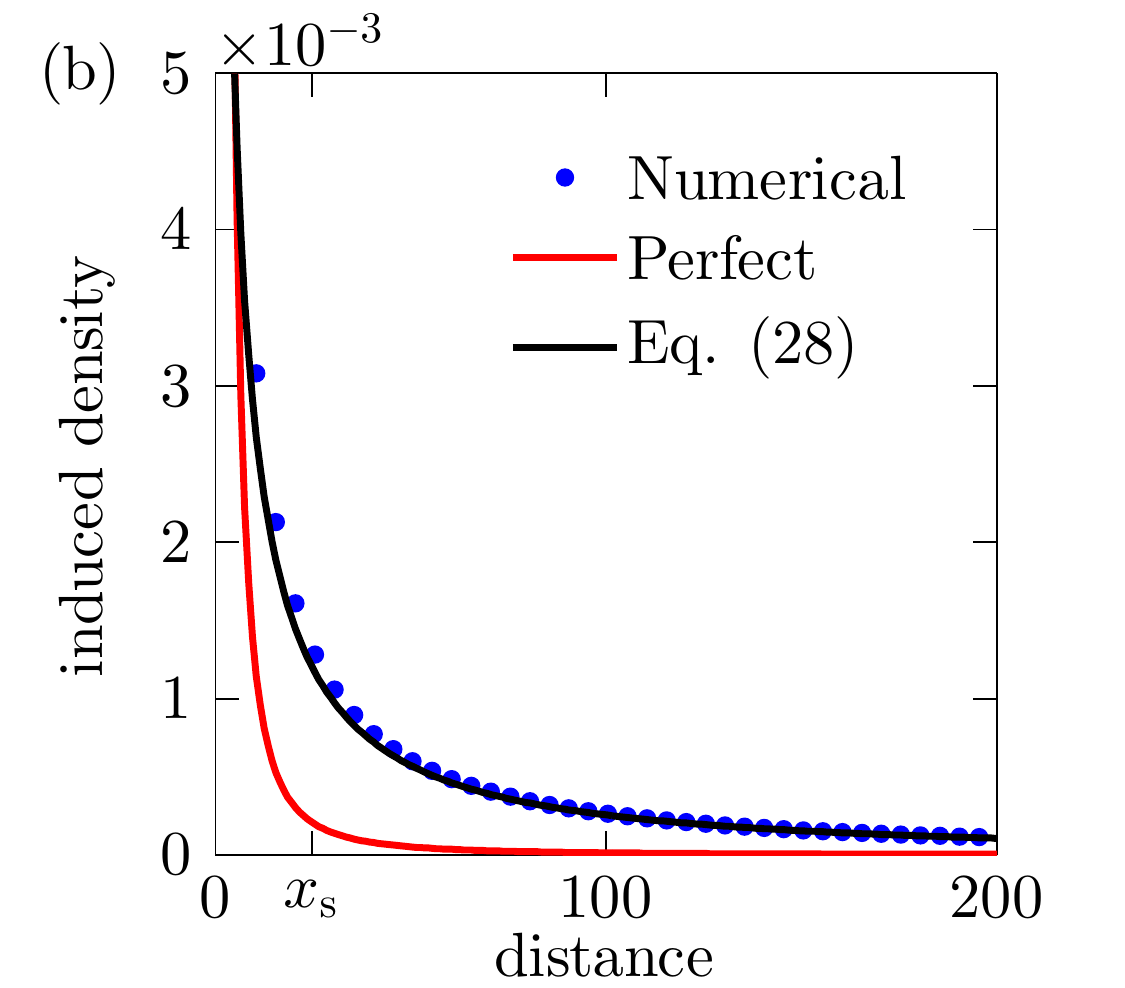}
		\includegraphics[height=2in]{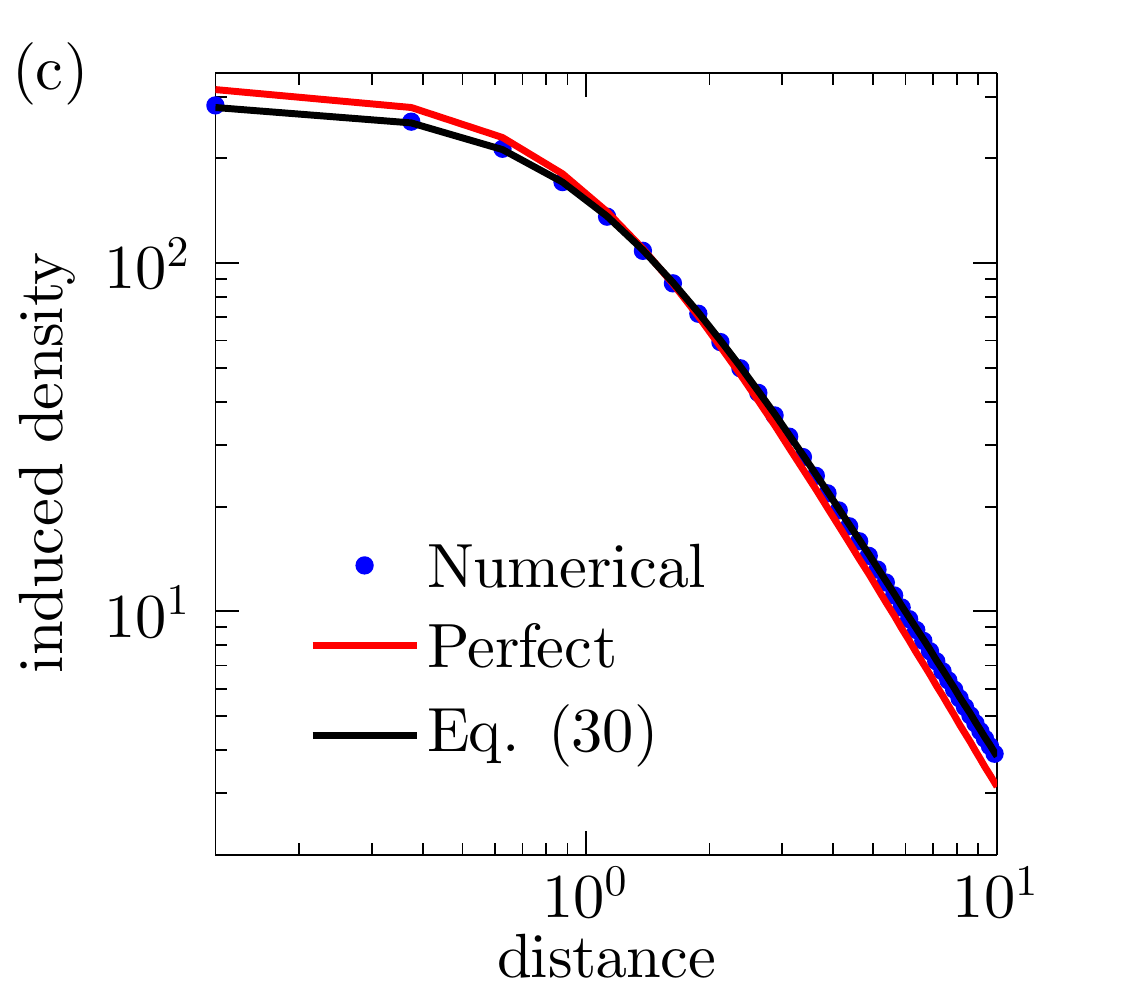}
	\end{center}
	\caption{(Color online) Comparison of induced electron density profile from the numerical solutions of the TFA equations (blue dots) and analytical formulas (black lines) in three screening regimes. 
	The red curves (labeled `Perfect') in each panel represent $n_\mathrm{ext}(x)$.
	Units for density $n$, distance $x$, and linear number density $\lambda$ are respectively $(4\pi^3\alpha^2 a^2)^{-1}$, $a$, and $(4\pi^3\alpha^2 a)^{-1}$. (a) Linear screening regime realized for $\lambda = 0.1$ and $n_\infty = 4$. 
	The induced density is much smaller than $n_\infty$.
	(b) Nonlinear regime realized for $\lambda=0.4$ and $n_\infty=0$. 
	Density profile characterized by screening length $x_s\approx 25$.
	(c) Near-perfect screening for $\lambda=1000$ and $n_\infty=0$.
	Note the double logarithmic scale.
	}
	\label{fig:simulation}
\end{figure*}

Linear screening of linelike charges by doped graphene has been studied in previous literature.~\cite{Radchenko2013ecl}
We include a brief summary of the relevant results for later comparison with our nonlinear response theory.
Linear screening is realized when the external charge is not too high or when graphene doping level is not too low.
The quantitative criterion $\tilde{\lambda} \ll 1$
is derived below.
Within the linear-response theory, the induced electron density  is given by
\begin{equation}
	n_\mathrm{ind}(x)
	= \int\limits_{-\infty}^{\infty}\frac{{d}q}{2\pi}
	\left[1-\frac{1}{\epsilon(q)}\right]
	\lambda e^{-a|q| + i q x}\, ,
	\label{eqn:n_ind_x}
\end{equation}
where the term $\lambda e^{-a |q|}$ is the Fourier transform of the effective external charge $n_{\mathrm{ext}}(x)$ [Eq.~\eqref{eqn:n_ext}] at momentum $q$.
As we are primarily concerned with distances $x\gg k_\infty^{-1}$ where the TFA is valid, the dielectric function of graphene can be approximated by~\cite{Kotov2012eei, Ando2002dca, Wunsch2006dpo}
\begin{equation}
	\epsilon(q) = 1 + \frac{q_\mathrm{TF}}{|q|}\,,
	\quad
	q_\mathrm{TF} = 4\alpha |k_\infty|\,.
	\label{eqn:epsilon_q}	
\end{equation}
A particularly simple analytical expression for $n_\mathrm{ind}(x)$ can be derived in the `asymptotic' regime, cf.~Fig.~\ref{fig:regimes}.
Carrying out the integration in Eq.~\eqref{eqn:n_ind_x}
by the steepest-descent method,
one finds the leading-order approximation for the induced density to be~\cite{Radchenko2013ecl}
\begin{equation}
	n_\mathrm{ind}(x) \simeq \frac{b}{x^2}\,,
	\quad b = \frac{\lambda}{\pi} \left(a + q_\mathrm{TF}^{-1}\right).
	\label{eqn:n_ind_ln}
\end{equation}
(In contrast, $n_\mathrm{ind} \propto x^{-3}$ for a pointlike charge perturbation.~\cite{Wunsch2006dpo})
Note that the coefficient $b$ in Eq.~\eqref{eqn:n_ind_ln} is much larger than $\lambda a / \pi$ under the assumed condition~\eqref{eqn:doping},
so that $n_\mathrm{ind}(x) \gg n_\mathrm{ext}(x)$.
Such an `overshoot' is typical for screening of localized perturbations in metals.
Metallic systems possess the overall charge neutrality.
However, at distances shorter than the local screening length from the perturbation screening is necessarily weak.
Therefore, there is a missing charge at short distances,
which has to be compensated at large $x$.
However, the electric field is \textit{not} overscreened:
it is of the same sign as the external one but reduced in magnitude.

The analytical Eq.~\eqref{eqn:n_ind_ln} agrees well with
our numerical simulations shown in Fig.~\ref{fig:simulation}(a).
For these simulations we used previously developed codes~\cite{Zhang2008nsb} with suitable modifications.
In brief, the electron density $n(x)$ to be found was defined on a grid of $x$ with periodic boundary conditions.
The solution was obtained by minimizing the total energy of the system (kinetic plus electrostatic) within the TFA using
standard technical computing software.~\cite{MATLAB}

Refinement of Eq.~\eqref{eqn:epsilon_q} can be obtained through the
random-phase approximation (RPA).
Within the RPA, the dielectric function of graphene coincides
with Eq.~\eqref{eqn:epsilon_q} at $0 < |q| < 2\, |k_\infty|$
but at $|q| > 2\, |k_\infty|$ it is given by a different formula~\cite{Ando2002dca, Wunsch2006dpo}
\begin{equation}
\begin{split}
\epsilon(q) &= 1 + \frac{q_\mathrm{TF}}{|q|}
- \frac{q_\mathrm{TF}}{2 |q|}\, \sqrt{1 - \left|\frac{2 k_\infty}{q}\right|^2}\\
&+ \alpha \cos^{-1}\left|\frac{2 k_\infty}{q}\right|\,,
\quad |q| > 2 |k_\infty|\,.
\end{split}
\label{eqn:epsilon_q_RPA}
\end{equation}
Notably, $\epsilon \simeq  1 + \pi \alpha / 2$ becomes doping independent at $|q| \gg 2\, |k_\infty|$ where
the response is dominated by virtual interband transitions.
(For corrections to the last result beyond RPA,
see Refs.~\onlinecite{Kotov2012eei, Sodemann2012icp}.)

Substituting Eq.~\eqref{eqn:epsilon_q_RPA} in Eq.~\eqref{eqn:n_ind_x}
and using contour integration techniques to evaluate the integral, we find
the RPA correction to $n_\mathrm{ind}$:
\begin{align}
\Delta n_\mathrm{ind}^\mathrm{RPA}(x) &\simeq
b_1\,\frac{\cos \left(2\, |k_\infty x| + \frac{\pi}{4}\right)}
        {|k_\infty  x|^{5 / 2}}\,,
\quad |k_\infty x| \gg 1\,,
\label{eqn:Delta_n_RPA}
\\
b_1 &= -\frac{\lambda}{2 \sqrt{\pi}}\,
  \frac{\alpha |k_\infty|}{(1 + 2\alpha)^2}\,,
\end{align}
which is a particular case of the Friedel oscillations.~\cite{Ashcroft1976ssp}
At intermediate distances, Eq.~\eqref{eqn:n_ind_x} yields~\cite{Radchenko2013ecl}
\begin{equation}
n_\mathrm{ind}(x) \simeq \frac{\lambda q_\mathrm{TF}}{\pi}\,
 \log \frac{0.561}{|q_\mathrm{TF} x|}\,,
 \quad
 1 \ll |k_\infty x| \ll \alpha^{-1}\,.
\label{eqn:n_ind_ln_weak}
\end{equation}
Finally, let us estimate the region of validity of the linear-response theory.
This theory applies if the induced carrier density is smaller than the original one, $n_\mathrm{ind} \ll n_\infty$ or,
equivalently, if the local Fermi momentum,
\begin{equation}
k_F(x) = \mathrm{sgn}\bigl(n(x)\bigr)
         |\pi n(x)|^{1 / 2}
\label{eqn:k_F}
\end{equation}
is perturbed slightly compared to $k_\infty$ [Eq.~\eqref{eqn:k_infty}].
Naively, one may require the condition $[k_F(x) - k_\infty] / k_\infty \ll 1$ to hold at all
distances of interest, $x \gg a$.
In fact, the validity region is wider because 
the nonlinearity affects only the response at momenta $q < 2\, |k_\infty|$.
(As mentioned above, the response at $q > 2\, |k_\infty|$ is essentially doping-independent.)
Therefore, the smallest number we should use in the argument of $n_\mathrm{ind}(x)$ for our estimate is $x \sim |k_\infty|^{-1}$.
From Eqs.~\eqref{eqn:n_ind_def}, \eqref{eqn:n_ind_ln_weak},
and \eqref{eqn:k_F} we get
\begin{equation}
\max \left[\frac{k_F(x)}{k_\infty} - 1\right]
\simeq
\frac{2 \alpha \lambda}{k_\infty}\,
 \log \left(\frac{1}{\alpha}\right)\,.
\label{eqn:k_F_change}
\end{equation}
Neglecting the logarithmic factor, which is never large in practice,
we arrive at $|\tilde{\lambda}| \ll 1$ as the criterion of linear screening.

\section{Nonlinear response}
\label{sec:nonlinear}

In this Section we treat a more difficult case
$\tilde{\lambda} \gg 1$ where screening is nonlinear.
Our approach to this problem is to solve the TFA equations analytically and numerically.
The first of these equations is 
\begin{equation}
	\mu(x) - e \Phi(x) = 0\,, 
	\label{eqn:TFA}
\end{equation}
where
\begin{equation}
\mu(x) = \hbar v k_F(x)
\label{eqn:g_dispersion}
\end{equation}
is the local chemical potential of graphene
(assuming the linear Dirac dispersion of quasiparticles).
The second equation links the total charge density and the electrostatic potential,
\begin{equation}
\Phi(x) = \frac{2 e}{\kappa} \int d x' \log |x - x'|
 [n_\mathrm{ind}(x') - n_\mathrm{ext}(x')]\,.
\label{eqn:Phi_III}
\end{equation}
This relation can be inverted by exploiting techniques from the theory of analytic functions:
\begin{equation}
	n_\mathrm{ind}(x) - n_\mathrm{ext}(x)
	= \frac{\kappa}{\pi^2 e}\, \mathcal{P}
	\int\limits_{0}^{\infty} \frac{x' {d}x'}{x'^2-x^2}\,
	\frac{d \Phi}{d x'}\,,
	\label{eqn:n_ind_KK}
\end{equation}
where $\mathcal{P}$ stands for the Cauchy principal value.
If desired, Eqs.~\eqref{eqn:TFA}--\eqref{eqn:n_ind_KK} can be combined into a single nonlinear integral equation for $n(x)$.
The TFA is valid if~~\cite{Landau1981qmn}
\begin{equation}
	\frac{{d}}{{d}x}k_F^{-1}(x) \ll 1\,.
	\label{eqn:TF_condition}
\end{equation}
The problem we want to solve can be separated into two parts,
depending on whether $n_\mathrm{ind}(x)$ is greater or smaller than $n_\infty$.
The latter situation occurs at large $x$,
where we expect screening behavior akin to linear response.
Indeed, it is easily seen that $n_\mathrm{ind}(x)$ follows
Eq.~\eqref{eqn:n_ind_asymp}
provided the integral in Eq.~\eqref{eqn:n_ind_KK} is dominated by $x' \ll x$.
Invoking Eq.~\eqref{eqn:TFA}, we then obtain the asymptotic behavior
$\Phi \sim x^{-2}$
for $x \gg x_\infty$, where 
\begin{equation}
x_\infty = |k_\infty|^{-1} |\pi b|^{1 / 2}
\label{eqn:x_infty}
\end{equation}
and $b$ is to be determined below.

The analytical form of $n_\mathrm{ind}(x)$ at $x \ll x_\infty$ is not immediately obvious.
Fortunately, were are able to find (by trial and error) the following approximate solution:
\begin{equation}
	n_\mathrm{ind}(x) \simeq \frac{1}{4 \pi^3 \alpha^2}\,
	\frac{1}{(x-x_\mathrm{s})^2}\,
	\log^2 \left(\frac{x}{x_\mathrm{s}}\right),
	\label{eqn:n_ind_nl}
\end{equation}
which is characterized by a nonlinear screening length $x_s$.
This length is found from the argument that at small $x$ the external field is nearly unscreened, so that the total and the external potentials differ only by some constant: $\Phi(x) \simeq \Phi_\mathrm{ext}(x) + \mathrm{const} = -2\lambda(e^2/\kappa)\log x + \mathrm{const}$ [Eq.~\eqref{eqn:Phi_ext_II}].
Comparing with Eq.~\eqref{eqn:n_ind_nl} at $x\ll x_s$, we get
\begin{equation}
	x_\mathrm{s}=\frac{1}{4\pi\alpha^2\lambda}\, .
\label{eqn:x_s}
\end{equation}
From Eqs.~\eqref{eqn:TF_condition} and \eqref{eqn:x_s}
we see that Eq.~\eqref{eqn:n_ind_nl} is valid at $x \gg 2\pi\alpha x_s$.
At smaller $x$ the density and the Fermi momentum presumably tend to a finite maximum, i.e,
\begin{equation}
\max k_F(x) \simeq k_F(2\pi \alpha x_s)
 \simeq 2 \alpha \lambda
 \log\left(\frac{1}{\alpha}\right)\,.
\label{eqn:k_F_0}
\end{equation}
The last logarithmic factor is valid if $\alpha \ll 1$;
otherwise,
it should be replaced by a number of the order of unity.
Equation~\eqref{eqn:k_F_0} is consistent with Eq.~\eqref{eqn:k_F_change} for the linear regime.
Hence, in both regimes the nonlinearity parameter $\tilde{\lambda}$
[Eq.~\eqref{eqn:tilde_lambda}] has the physical meaning of the maximum relative change in the Fermi momentum $(k_F - k_\infty) / k_\infty$ caused by the perturbation.

To fix the so far undetermined coefficient $b$,
we require a smooth matching of Eq.~\eqref{eqn:n_ind_asymp} and Eq.~\eqref{eqn:n_ind_nl}, which yields
\begin{equation}
b \simeq \frac{1}{4\pi^3\alpha^2}\,
\log^2 |\tilde{\lambda}|\,,
\quad
x_\infty \simeq 
\frac{1}{2\pi\alpha k_\infty}\,
	\log |\tilde{\lambda}|\,.
\label{eqn:x_infty}
\end{equation}

More accurate expression for $n_\mathrm{ind}(x)$ can be obtained by iterations using Eq.~\eqref{eqn:n_ind_nl} as the input to the TFA equations.
Namely, substituting it into Eq.~\eqref{eqn:TFA}, we can get $\Phi(x)$,
which we can then insert into Eq.~\eqref{eqn:n_ind_KK} to obtain an improved approximation for $n_\mathrm{ind}(x)$. 
The first iteration yields
\begin{equation}
	\begin{split}
	 n_\mathrm{ind}^{(1)}(x) &=  \frac{1}{4\pi^3\alpha^2}\Bigg[\frac{\pi^2}{2}\frac{1}{(x+x_\mathrm{s})^2} \\
	& +\log^2\left(\frac{x}{x_s}\right)\frac{x^2+x_s^2}{(x^2-x_s^2)^2}-2\frac{\log(x/x_\mathrm{s})}{x^2-x_\mathrm{s}^2}\Bigg],
	\end{split}
\label{eqn:n_ind_nl_corr}
\end{equation}
which demonstrates a good agreement with our numerical simulations, see Fig.~\ref{fig:simulation}(b).

For completeness, we consider the case of
\begin{equation}
\lambda > \frac{1}{4\pi\alpha^2 a}\,,
\end{equation} 
where nonlinear screening affects
distances shorter than our cutoff length $a$.
At such large $\lambda$,
a highly doped region appears at $x < a$,
where screening is nearly perfect, i.e.,
where the difference between $n_\mathrm{ind}(x)$ and $n_\mathrm{ext}(x)$ is relatively small and the system is able to maintain the local charge neutrality.
The solution for $n_\mathrm{ind}(x)$ can again be obtained by iterations.
The input to Eq.~\eqref{eqn:n_ind_KK} is now
$n_\mathrm{ind}(x) = n_\mathrm{ext}(x)$.
The first correction is
\begin{equation}
n_\mathrm{ind}(x) - n_\mathrm{ext}(x)
 \simeq \frac{\sqrt{\lambda a}}{\pi^2\alpha}\left[
\frac{x \sinh^{-1} \frac{x}{a}}{(x^2 + a^2)^\frac32}
 - \frac{1}{x^2 + a^2}
 \right].
\label{eqn:n_ind_np}
\end{equation}
Its range of validity $x < x_p$ can be estimated from the requirement that at $x = x_p$ this correction is no longer small,
which implies
\begin{equation}
\log (x_p / a) \sim \pi\alpha\sqrt{\lambda a}\,.
\label{eqn:x_p}
\end{equation}
Furthermore, one can check that at $x = x_p$
Eq.~\eqref{eqn:n_ind_np}
matches by the order of magnitude with $n_\mathrm{ind}(x)$ given by Eq.~\eqref{eqn:n_ind_nl} with $x_s$ set to $a$.
This leads to us conclude that at $x > x_p$ Eq.~\eqref{eqn:n_ind_nl} and then at $x > x_\infty$ Eq.~\eqref{eqn:n_ind_asymp} must still hold with $x_s \sim a$.
Because of the exponentially large magnitude of $x_p$,
we could verify numerically only the $x < x_p$ case.
Comparison of simulations with Eq.~\eqref{eqn:n_ind_np} in Fig.~\ref{fig:simulation}(c) indeed shows a good agreement.

If $x_p > \sqrt{\lambda a} / k_\infty$, which corresponds to
\begin{equation}
\lambda > \frac{1}{\pi^2\alpha^2 a}
          \log^2 \left(\frac{1}{\alpha a k_\infty}\right)\, ,
\end{equation}
the screening should be nearly `perfect' for all $x > a$, cf.~Fig.~\ref{fig:regimes}.

When $\lambda$ and $n_\infty$ have opposite signs, twin $p$-$n$ junctions appear at $|x| \sim x_\infty$. 
The density profile near a graphene $p$-$n$ junction has been
studied in Ref.~\onlinecite{Zhang2008nsb}.
These results should more or less carry over to the present problem, so
they will not be repeated.
In the following Sections we focus on computing
the effect of nonlinear screening
on transport and capacitance characteristics of the system.

\section{Quantum Capacitance}
\label{sec:capacitance}

The charge density $\lambda$ is a natural parameter for graphene grain boundaries [Fig.~\ref{fig:schematic}(b)].
In contrast, linelike perturbation created by means of narrow gates [Fig.~\ref{fig:schematic}(a)] are controlled by the voltage $V$ with respect to graphene
while $\lambda$ has to be found by integrating the differential gate capacitance
\begin{equation}
C(V) = e\, \frac{d \lambda}{d V}\,.
\label{eqn:C}
\end{equation}
It is important that $V$ is not simply equal to the electrostatic potential difference $\Delta\Phi$ between the gate and the graphene sample.
It has another contribution from the graphene chemical potential:
\begin{equation}
V = \Delta\Phi + (\mu / e)\,.
\label{eqn:V}
\end{equation}
As a result, 
the differential gate capacitance $C$
has two components, the classical (or geometrical) one $C_g$ and the so-called quantum one $C_q$, which add in series:
\begin{equation}
C^{-1} = C_g^{-1} + C_q^{-1}.
\end{equation}
Our goal in this Section is to derive the quantum capacitance of a device with an ultranarrow top gate, Fig.~\ref{fig:schematic}(a).
However, it is useful to review the conventional planar-gate geometry first.
Here the classical capacitance per unit area $C_g = \kappa / (4\pi d)$ is inversely proportional to the separation $d$ between graphene and the planar gate.
In turn, the quantum capacitance is proportional to the thermodynamical density of states, $C_q = e^2 (d n / d \mu)$.
This quantity can also be written in terms of the inverse Thomas-Fermi
screening length $q_\mathrm{TF}$:
\begin{equation}
C_q = \frac{\kappa}{2\pi}\, q_\mathrm{TF}\,.
\label{eqn:C_q_pp}
\end{equation}
The net result of having a finite screening length is equivalent to replacing the physical gate-graphene separation by an effective one:
\begin{equation}
d_\mathrm{eff} = d + \frac12\, q_\mathrm{TF}^{-1}\,.
\label{eqn:d_eff}
\end{equation}
Although in this article we use the free-fermion approximation [Eq.~\eqref{eqn:epsilon_q}] for
the linear-response screening length $q_\mathrm{TF}^{-1}$,
in reality it is modified by many-body interactions and disorder (see, e.g., Refs.~\onlinecite{Fogler2004nsp, Allison2006tds}).
Recently quantum capacitance measurements have been used to probe such effects of graphene.~\cite{Yu2013ipg, Basov2014cgs}

\begin{figure}
	\includegraphics[height=2in]{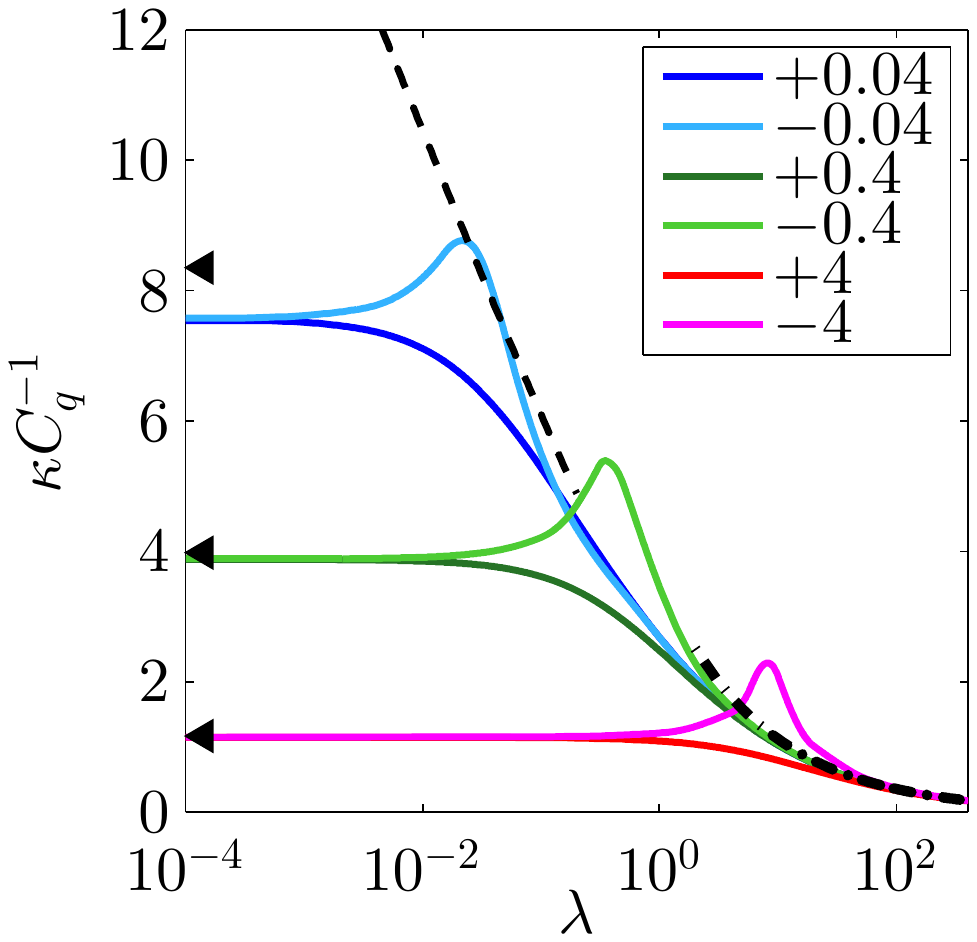}	
	\caption{(Color online) Inverse quantum capacitance as a function of the gate charge.
	The curves are from numerical calculations
	for $k_\infty$ specified in the legend.
	The units of $\lambda$ and $k_\infty$ are $(4\pi^3\alpha^2a)^{-1}$ and $(\pi^3\alpha a)^{-1}$.
	The triangles, the dashed line, and the dash-dotted line correspond, respectively, to the asymptotic limits of the linear, nonlinear, and near-perfect screening, Eqs.~\eqref{eqn:cq_ln}, \eqref{eqn:cq_nl}, and
	\eqref{eqn:cq_cp}.
	The difference between the curves and the triangles at smaller $k_\infty$ are due to finite-size effects in the simulation.
	The peaks of the $k_\infty < 0$ curves signal the formation of the twin $p$-$n$ junctions.
	}
	\label{fig:capacitance}
\end{figure}

If the gate is now a long metallic string or radius $l \ll a$,
the capacitances per unit length are relevant.
In order to derive $C^{-1}$ we start with
a general expression for the electrostatic potential difference between the string and the graphene sample,
\begin{equation}
	\begin{split}
	\Delta \Phi(x) &=\frac{e\lambda}{\kappa}
	 \log\left(\frac{x^2 + a^2}{l^2}\right)\\
	&+\frac{e}{\kappa} \int_{-\infty}^{\infty} d x' n_\mathrm{ind}(x')
	\log\frac{x'^2 + a^2}{(x - x')^2}\,,
	\end{split}
\label{eqn:Delta_Phi}
\end{equation}
which follows from Eqs.~\eqref{eqn:Phi_ext} and \eqref{eqn:Phi_III}.
If we set $n_\mathrm{ind} = n_\mathrm{ext}$,
we obtain
$\Delta \Phi = ({e\lambda} / {\kappa}) \log(4a^2 / l^2) = \mathrm{const}$.
Hence, the geometric capacitance of the system is
\begin{equation}
C_g^{-1} = \frac{\Delta \Phi}{e \lambda}
= \frac{2}{\kappa} \log\left(\frac{2a}{l}\right),
\label{eqn:C_g}
\end{equation}
which can be alternatively derived by the method of images.
Next, combining Eqs.~\eqref{eqn:TFA}, \eqref{eqn:Phi_III}, \eqref{eqn:C}, \eqref{eqn:V}, \eqref{eqn:Delta_Phi},
and subtracting $C_g^{-1}$,
we find
%
\begin{equation}
C_q^{-1} = \frac{1}{e\lambda}
\int_{-\infty}^{\infty} d x\,
n_\mathrm{ext}(x)
\frac{\partial}{\partial\lambda}\, \Phi(x)\,.
\label{eqn:cq_II}
\end{equation}
We can now use this expression for analytical and numerical calculations.
Our analytical formulas for positive $\tilde{\lambda} \equiv \lambda / (\alpha k_\infty)$ are as follows:
\begin{subnumcases}
{\frac{\kappa}{2}\, C_q^{-1}\simeq}
\label{eqn:cq_ln}
 -e^{2a q_\mathrm{TF}}\text{Ei}(-2a q_\mathrm{TF})\,,
  & $\tilde{\lambda}\ll 1$,\\
\label{eqn:cq_nl}
 -\log (4 \pi\alpha^2 a \lambda)\,,
  & $\tilde{\lambda}\gg 1$,
\end{subnumcases}
where $\text{Ei}(z)$ is the exponential integral.
These equations describe, respectively, the linear and the `strong' regimes of Fig.~\ref{fig:regimes}.
They can be commonly written as
\begin{equation}
C_q^{-1}\simeq \frac2\kappa\log\left(\frac{x_\mathrm{sc}}{2 a}\right),
\label{eqn:C_q_x_sc}
\end{equation}
where $x_\mathrm{sc}$ is equal to (in the same order) 
$q_\mathrm{TF}^{-1}$ and $x_s$.
We conclude that the total capacitance can be modeled after the geometric one,
\begin{equation}
C^{-1} = \frac2\kappa \log\left(\frac{2 a_\mathrm{eff}}{l}\right)\,
\end{equation}
with the effective gate-graphene separation
\begin{equation}
a_\mathrm{eff} = a + \frac{x_\mathrm{sc}}{2}\,,
\label{eqn:a_eff}
\end{equation}
where $a$ on the right-hand side was added by hand to recover the result $a_\mathrm{eff} = a$ expected for a perfect metal, $x_\mathrm{sc} = 0$.
The similarity of Eqs.~\eqref{eqn:d_eff} and \eqref{eqn:a_eff}
illustrates once again that quantum capacitance is a measure of the screening length of a system.
Whereas in the planar-gate geometry this length is formally divergent for undoped graphene,
for the linelike gate the divergence is regularized by nonlinearity.
Numerically, we find that $C_q^{-1}$ as a function of $\lambda$ approaches a universal envelope curve~\eqref{eqn:cq_nl} shown by the dashed line in Fig.~\ref{fig:capacitance}.

For completeness, we consider the `perfect' screening regime where
\begin{equation}
e\Phi(x) \simeq n_\mathrm{ext}(x)\, \frac{d\mu}{d n}
\label{eqn:Phi_perfect}
\end{equation}
for all relevant $x$, so that Eq.~\eqref{eqn:cq_II} becomes
\begin{equation}
C_q^{-1} \simeq \frac{1}{e^2} 
\int_{-\infty}^{\infty}
\frac{d x}{\lambda^2}\,
n_\mathrm{ext}^2(x)
\frac{d \mu}{d n}\,.
\label{eqn:cq_np_apprx}
\end{equation}
The inverse thermodynamic density of states ${d\mu} / {d n}$ in Eqs.~\eqref{eqn:Phi_perfect} and \eqref{eqn:cq_np_apprx}
is to be evaluated at $n = n_\infty + n_\mathrm{ext}(x)$.
For $n_\infty = 0$ where the integrand scales as $|\lambda|^{1 / 2}$, we find the analytical result
\begin{equation}
\frac{\kappa}{2}\, C_q^{-1} \simeq
 \frac{1}{2\pi\alpha\sqrt{\lambda a}}\,,
\quad
\lambda > \dfrac{1}{\alpha^2 a}\,,
\label{eqn:cq_cp}
\end{equation}
which agrees with our numerical simulations (the dashed-dotted curve in Fig.~\ref{fig:capacitance}).

If $\tilde{\lambda}$ is negative, i.e.,
if $\lambda$ and $k_\infty$ have opposite signs, the twin $p$-$n$ junctions form at some $\lambda$ which can be estimated from Eq.~\eqref{eqn:k_F_change} setting $k_F(0)$ to zero.
This event --- the onset of the ambipolar regime --- is marked by a maximum in $\kappa C_q^{-1}$,
which is absent in the unipolar trace for the same $|k_\infty|$, see Fig~\ref{fig:capacitance}.
From dimensional arguments, as $k_\infty$ approaches zero,
the height of the maximum in $\kappa C_q^{-1}$ measured with respect to its plateau at $\lambda = 0$ should approach a universal number.
Figure~\ref{fig:capacitance} suggests that number is about $1.5$.
However, such dimensional arguments assume the TFA is valid, which,
similar to the case of a single $p$-$n$ junction,~\cite{Zhang2008nsb} is the case at small $\alpha$.
To treat a more typical case $\alpha \sim 1$ one needs 
to go beyond the TFA,
which may be a problem for future research.

\section{Conductance}
\label{sec:transport}

Charged linelike defects are known to significantly influence electron transport in graphene.
Grain boundaries strongly reduce the sheet conductivity of large-area graphene,~\cite{Tsen2012tet}
while bipolar junctions created by nanowire gates cause conductance oscillations.~\cite{Young2009qik}
In this section we find expressions for the graphene conductance $G$ relevant for both situations.

We consider a scattering problem for a massless Dirac particle with initial momentum $\mathbf{k} = |k_\infty| (\cos\theta, \sin\theta)$ subject to the potential perturbation
\begin{equation}
-e\Phi(x) = \mu(n_\infty) - \mu(n)
 = \hbar v [k_\infty - k_F(x)]\,.
\label{eqn:ePhi}
\end{equation}
The intermediate equation follows from the TFA, Eq.~\eqref{eqn:TFA}.
For a weak potential, we can apply the standard perturbation theory to the massless Dirac equation
to obtain the reflection coefficient
\begin{equation}
r(\theta) = i \tan\theta\,
\int\limits_{-\infty}^{\infty} e^{2 i k_x x} k_F(x) d x\,,
\label{eqn:Born}
\end{equation}
which is similar to the first Born approximation formula for the Schr\"odinger equation.~\cite{Landau1981qmn}
The region of validity of this formula can be extended beyond the perturbative regime if in the argument of the exponential we replace $k_x$ by $k_F(0) \cos\theta$, the local momentum at the $x = 0$ point where the scattering potential is the `most' nonanalytic.~\cite{Landau1981qmn}
However, this is permissible only if $k_F(x)$ is real at all $x$,
i.e., if all points on the quasiparticle path are classically allowed (no quantum tunneling occurs).

\begin{figure}[tb]
\includegraphics[height=2in]{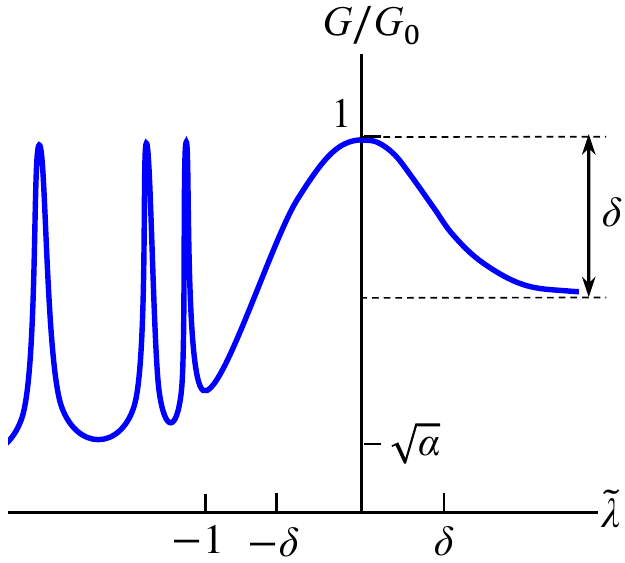}
\caption{Reduced conductance $G / G_0$ as a function of $\tilde{\lambda}$ (schematically).
A monotonic decrease is expected in the unipolar case, $\tilde{\lambda} > 0$, while Fabry-P\`{e}rot oscillations with a period determined by Eq.~\eqref{eqn:phi} should dominate for negative and large $\tilde{\lambda}$
where the $p$-$n$ junctions form.
Parameter $\delta$ is defined in the text.
}
\label{fig:conductance}
\end{figure}

The conductance $G$ is found by summing the transmittances $T(\theta)$
of all the $k_y = k_\infty \sin \theta$ channels:
\begin{equation}
G = \frac{4e^2}{h} \sum_{k_y} T(\theta)\,,
\quad T(\theta) = 1 - |r(\theta)|^2\,.
\end{equation}
In the absence of scattering, $\lambda = 0$, the conductance is
\begin{equation}
G_0 = \frac{4e^2}{h}\, \frac{|k_\infty| W}{\pi}\,,
\end{equation}
where $W$ is the width of the graphene sheet.
Assuming $W \gg k_\infty^{-1}$, we can compute $G$ at $\lambda \neq 0$
by replacing the summation with the integration over $\theta$.
However, $|r(\theta)|$ diverges as $|\theta|$ tends to $\pi / 2$, so that the first Born approximation cannot be used.
In fact, the absolute value of the exact reflection coefficient must be approaching unity instead of diverging.
We account for this by cutting off the integration limits at
$\bar{\theta}$ where $|r(\bar{\theta})| \sim 1$:
\begin{equation}
G = G_0 \int\limits_0^{\bar{\theta}}
 T(\theta) \cos\theta d\theta\,.
\end{equation}
For the linear screening regime, $\tilde{\lambda} \ll 1$ [Eq.~\eqref{eqn:tilde_lambda}],
we obtain
\begin{equation}
\frac{G_0 - G}{G_0}\simeq
\begin{cases}
  \phantom{\Bigg|} \dfrac{\tilde{\lambda}^2}{\alpha^2}\,
     \log^2 |\tilde{\lambda}|\,,
      & |\tilde{\lambda}| \ll \alpha^2\,,\\
\dfrac{\pi}
      {\bigl|\tilde{\lambda}^{-1} + \delta^{-1} \bigr|}\,,
       &\alpha^2 \ll |\tilde{\lambda}| \ll \delta\,.
\end{cases}
\end{equation}
Accordingly, the conductance $G$ as a function of $\tilde{\lambda}$ exhibits
an asymmetric maximum at $\tilde{\lambda} = 0$ drawn
schematically in Fig.~\ref{fig:conductance}.
The asymmetry becomes pronounced when $\tilde{\lambda}$ approaches $\delta = 1 / (2\,\log \alpha^{-1})$,
where the screening begins to cross over into the nonlinear regime.
If $\tilde{\lambda}$ is positive,
$k_F(x)$ becomes essentially independent of $k_\infty$.
Using Eqs.~\eqref{eqn:n_ind_nl} and \eqref{eqn:k_F_0}
and changing the integration variable $x \to \lambda x$ in Eq.~\eqref{eqn:Born} one can show that $r(\theta)$ does not depend on $\lambda$ any more in this regime.
Actually, this is clear from dimensional argument:
since $r(\theta)$ is dimensionless, it may not depend on $\lambda$, which has the units of inverse length.
This implies that $G$ ceases to decrease with $\tilde{\lambda}$, leveling at a plateau,
$(G_0 - G) / G_0 \simeq \pi \delta$.
If $\tilde{\lambda}$ is negative, the
situation is quite different.
The scattering potential is repulsive.
At large $|\tilde{\lambda}|$ it causes the twin $p$-$n$ junctions to appear at $|x| \sim x_\infty$
[Eq.~\eqref{eqn:x_infty}],
which act as tunneling barriers.
Here even the modified Born approximation fails completely.
The transmittance $T_{pn}(\theta)$ 
of each $p$-$n$ junction is given instead by~\cite{Cheianov2006std}
\begin{equation}
T_{pn}(\theta) = \exp\left(-\frac{\pi\hbar v k_\infty^2}{F}\,
 \sin^2\theta\right),
\end{equation} 
where 
\begin{equation}
F = 2.5\, \hbar v \alpha^{1/3}|n'|^{2/3}
\end{equation} 
is the electric field at the junction,~\cite{Zhang2008nsb}
with $n'$ being the density gradient.
Combining these equation, we obtain
\begin{equation}
T_{pn}(\theta) = \exp\left(-\frac{b_2}{\alpha}\,
 \sin^2\theta\right)\,,
\quad b_2 \sim \log^{2 / 3} |\tilde \lambda|\,.
\label{eqn:t_pn_II}
\end{equation} 
Multiple reflections of the quasiparticles in the region between the $p$-$n$ junctions lead to the conductance oscillations and resonances.
The net transmittance is given approximately by the Fabry-P\'{e}rot-like formula~\cite{Young2009qik}
\begin{equation}
T(\theta) = \left|
\frac{T_{pn}(\theta)}{1 - [1 - T_{pn}(\theta)]
e^{i \phi}}
\right|^2\,,
\end{equation}
where $\phi$ is the phase acquired by a quasiparticle after one roundtrip between the junctions:
\begin{equation}
\phi \simeq
 2\int\limits_{-x_\infty}^{x_\infty} d x\, k_F(x)
 \simeq \frac{1}{\pi\alpha}\,
 \log^2 |\tilde{\lambda}|\,.
\label{eqn:phi}
\end{equation}
The last estimate is obtained using Eqs.~\eqref{eqn:n_ind_nl} and \eqref{eqn:x_infty}.
Conductance minima arise at $\phi = (2 m + 1)\pi$, where $m$ is an integer.
They have the magnitude $G_\mathrm{min} / G_0 \sim \sqrt{\alpha / b_2}$
because the transmittance of each junction is appreciable only at small angles~\cite{Cheianov2006std} $\theta < \sqrt{\alpha / b_2}$,
cf.~Eq.~\eqref{eqn:t_pn_II}.
Conductance maxima are found at $\phi_m = 2 m\pi$,
which correspond to $|\tilde{\lambda}_m| \sim \exp(\pi\sqrt{2 \alpha m}\,)$.
The widths of these maxima and the distance from one to the next increase exponentially as a function of $m$.
The heights of these maxima approach $G_0$,
which is a manifestation of the resonant tunneling phenomenon.
In practice, observation of the resonant tunneling requires
samples with the mean free path longer than the roundtrip distance $4 x_\infty$;
otherwise, the conductance is influenced by diffusive transport
between and across the $p$-$n$ junctions.~\cite{Fogler2008edg}
In previous experiments with nanowire-gated graphene devices,~\cite{Young2009qik} the conductance maxima were found to be significantly lower than $G_0$ and decreasing with the top gate voltage, i.e., $|\lambda|$, presumably due to disorder scattering. 

If the conductance can also be measured in the direction parallel to the gate,
we expect it
to show a dependence consisting of a smooth increase with $|\lambda|$ with superimposed small oscillations due to
quantization of the quasi-bound resonant modes.
This oscillating part would have the same period as the
Fabry-P\'{e}rot oscillations in the transverse conductance discussed
above.
At small gate voltages, both longitudinal and transverse
conductances are expected to show additional fine features related to
the analog of the Goos-H\"{a}nchen effect in graphene,~\cite{Beenakker2009qgh} which is a lateral displacement of a quasiparticle trajectory along the $p$-$n$ interface during reflection.
This effect can be included using a more accurate equation for $\phi$ that incorporates both the path length contribution expressed by Eq.~\eqref{eqn:phi} and the phase shift of the reflections at the $p$-$n$ interfaces.

\section{Discussion}
\label{sec:conclusion}

In this work we considered linelike charged perturbations on graphene
and derived analytical expressions for the induced density profiles in both linear and nonlinear screening regimes. 
These results were applied to the analysis of two types of electronic properties.
The first one is the quantum correction to the classical capacitance between the narrow gate and graphene as a function of the top gate voltage.
Measuring this quantity will be a direct way for observing the crossovers
among different screening regimes and testing our predictions.
For example, we showed that the divergence of the inverse quantum capacitance of undoped graphene predicted from the naive linear-response theory will be curbed by nonlinear screening effects.  
If the gate creates a strong repulsive potential for charge carriers in graphene, it can induce twin $p$-$n$ junctions.
Our calculations indicate that the onset of this ambipolar regime is signaled by a peak in the inverse quantum capacitance.

The second quantity we studied is the transverse electrical conductance of the system. Our predictions for the ambipolar regime, where the conductance oscillates
as a function of gate voltage, include formulas for the maxima, minima, and the oscillation period.
Our theory holds in the ballistic transport regime, which was difficult to probe
in earlier experiments on such systems.~\cite{Young2009qik}
We hope that modern higher-quality devices that utilize graphene encapsulated in boron nitride~\cite{Dean2010bns} and 
single-wall nanotube gates of smallest possible diameter,
would enable a systematic investigation of nonlinear screening and resonant tunneling phenomena we discussed.

Conductance of graphene with charged grain boundaries was previously studied analytically and numerically in Refs.~\onlinecite{Ferreira2011tpg, Radchenko2013ecl, Ihnatsenka2013eic}.
In Ref.~\onlinecite{Radchenko2013ecl}
transport properties were computed modeling charged grain boundaries as short line segments of length $W \ll k_\infty^{-1}$.
However, the screening was treated assuming that the boundaries are infinitely long, which seems to require the opposite inequality.
These incompatible assumptions make a direct comparison between our analytical results for the conductance difficult.
As for screening, only the linear regime was considered in Ref.~\onlinecite{Radchenko2013ecl}, and for this our results agree.

Our findings have further implication for experiments using novel scanned-probe techniques.
Linear charged defects in the form of grain boundaries have been shown to reflect surface plasmon polaritons~\cite{Fei2013epp,Schnell2014soh,Gerber2014prs} and induce photocurrent,~\cite{Woessner2015npn} both of which can be imaged with nanoscale resolution using scanning near-field optical microscopy.
Scanning tunneling microscopy is another avenue of approach to measure local density of states.~\cite{Koepke2013ase, Nemes-Incze2013esd}
We will apply our theory to interpretation of such measurements in a future work.~\cite{Ni2015xxx}

Finally, let us comment on the proposals~\cite{Katsnelson2007gnb} that graphene is a condensed-matter laboratory for exotic effects predicted in other fields of physics.
For example, electronic response of graphene to a pointlike charge has an interesting analogy to the atomic collapse of superheavy elements.~\cite{Pomeranchuk1945els, Zeldovich1972ess}
It has been shown~\cite{Katsnelson2006nsc, Novikov2007est, Shytov2007vps, Shytov2007acq, Fogler2007shc, Pereira2007cip, Kotov2012eei}
that in graphene
subcritical $Z < Z_c$ and supercritical $Z > Z_c$ charges produce qualitatively different behavior of the screened electrostatic potential at large distances from the perturbation,
the critical charge $Z_c$ being of the order of $1 / \alpha$.
Characteristic oscillations of the local density of states that appear in the supercritical case
have been recently detected experimentally.~\cite{Wang2013oac}
Phenomena similar to atomic collapse have been also studied theoretically in
the context of narrow-band gap semiconductors
and Weyl semimetals.~\cite{Kolomeisky2013fsc}
In turn, our problem of screening of a linelike charge perturbation in graphene have interesting
analogies in cosmology
(screening of a hypothetical cosmic string by vacuum polarization~\cite{Nascimento1999cvc}) and polyelectrolyte physics (Onsager-Manning condensation of counterions~\cite{Manning1969llc, Oosawa1968tel}).
Our results imply that in graphene nonlinear screening plays a greater role for linelike charges compared to the pointlike ones:
the former are \textit{always} supercritical, e.g.,
there is no threshold $\lambda$ for the appearance of Friedel oscillations.
Finally, our analytical formulas assume $\alpha \ll 1$,
which can be realized using a high-$\kappa$ dielectric substrate [Fig.~\ref{fig:schematic}(a)], such as SrTiO$_3$ [Refs.~\onlinecite{Couto2011ttg, DasSarma2012gst, Sachs2014fs, Saha2014utt}].
For such gate dielectrics it may be important to consider electric-field dependence of $\kappa$ in nonlinear screening regimes.~\cite{Fu2015ced}

\acknowledgments

The work has been supported by the Office of the Naval Research and by the University of California Office of the President.
We thank D.~N.~Basov, Z.~Fei, and G.~Ni for prior and ongoing collaborations that motivated this study and also A.~Ferreira and B.~I.~Shklovskii for useful discussions and comments on the manuscript.

\bibliography{line_charge}
\end{document}